# Two-dimensional Ferromagnetic van der Waals CrX$_3$ (X=Cl, Br, I) Monolayers with Enhanced Anisotropy and Curie Temperature

Feng Xue[1], Yusheng Hou[2], Zhe Wang[1], Ruqian Wu[2,*]

[1] State Key Laboratory of Surface Physics and Key Laboratory for Computational Physical Sciences (MOE) and Department of Physics, Fudan University, Shanghai, 200433, China
[2] Department of Physics and Astronomy, University of California, Irvine, CA 92697-4575, USA

## Abstract

Among the recently widely studied van der Waals layered magnets CrX$_3$ (X=Cl, Br, I), CrCl$_3$ monolayer (ML) is particularly puzzling as it is solely shown by experiments to have an in-plane magnetic easy axis and, furthermore, all of previous first-principles calculation results contradict this. Through systematical first-principles calculations, we unveil that its in-plane shape anisotropy that dominates over its weak perpendicular magnetocrystalline anisotropy is responsible for the in-plane magnetic easy axis of CrCl$_3$ ML. To tune the in-plane ferromagnetism of CrCl$_3$ ML into the desirable perpendicular one, we propose substituting Cr with isovalent tungsten (W). We find that CrWCl$_6$ has a strong perpendicular magnetic anisotropy and a high Curie temperature up to 76 K. Our work not only gives insight into understanding the two-dimensional ferromagnetism of van der Waals MLs but also sheds new light on engineering their performances for nanodevices.

Email: wur@uci.edu

# 1. Introduction

Ever since the discovery of grapheme [1], two-dimensional (2D) materials have attracted tremendous attention due to their unique properties and immense potential in nano device applications. Over the past decades, a variety of novel 2D materials have been successfully fabricated and extensively studied, including hexagonal boron nitride [2], silicone [3], transition metal dichalcogenides [4,5], and black phosphorous [6], to name a few. Recent discoveries of atomically thin ferromagnetic films such as $Cr_2Ge_2Te_6$ [7] and $CrI_3$ [8] added a new thrust in this realm as they allow integration of magnetic materials in multifunctional 2D heterostructures, essential for the design of various spintronic and topotronic devices [9-12]. Fundamentally, the long-range magnetic ordering in a 2D system is vulnerable to thermal fluctuations even at extremely low temperature, according to the Mermin-Wagner theorem [13] that was established from the isotropic Heisenberg model. A sizeable magnetic anisotropy energy which forces magnetic moments in a lattice to align in certain crystallographic direction known as easy axis becomes critical to stabilize the 2D magnetic ordering. Indeed, it was shown that the high Curie temperatures of $Cr_2Ge_2Te_6$ and $CrI_3$ monolayers (MLs) stem from their out-of-plane (perpendicular) magnetic anisotropy, that opens up an spin wave gap to resist the thermal agitation [7].

Chromium trihalides $CrX_3$ (X = Cl, Br, I) have the van der Waals (vdW) layered structure and share similar physical properties. Their monolayers have been targeted as the most promising 2D magnetic materials for fundamental studies and technological innovations. Theoretical studies suggest that the magnetic easy axes of all $CrX_3$ monolayers are out-of-plane [14,15] just like their bulks. While recent experiments [16-21] indeed confirm this conjecture for $CrBr_3$ and $CrI_3$ MLs, an in-plane magnetic easy axis was observed for the $CrCl_3$ ML [17,18]. As the perpendicular easy direction of magnetic MLs is essential for the sustainability of their ferromagnetic (FM) ordering and, subsequently, for their performance in heterostructures such as inducing spin splitting on the topological surface states [22], it is important to solve the puzzle for $CrCl_3$ ML. Furthermore, it is desired to explore effective strategies to control the magnetic anisotropy and to enhance the Curie temperature ($T_C$) of $CrX_3$ MLs.

In the present work, we performed systematical first principles calculations for $CrX_3$ MLs and their derivations. We found that the inclusion of magnetic shape anisotropy (MSA) is crucial for studies of these thin films and is the driving force for the in-plane magnetization of the $CrCl_3$ ML. By substituting part of Cr atoms with isovalent W atoms, the magnetic easy axis of $Cr_{1-x}W_xCl_3$ MLs can be tuned to the perpendicular direction with a large magnetic anisotropy energy (MAE) of 1.07 meV per magnetic atom. Moreover, the exchange interactions among magnetic atoms are also enhanced by W substitution, resulting in a higher $T_C$ up to 76 K. Therefore, our findings not only solve puzzles for the understanding of 2D magnetism in $CrX_3$ MLs, but also provide a viable strategy for optimizing their performance as needed for the development of spintronic materials and devices.

# 2. Theoretical approach

The density functional theory (DFT) calculations were performed using the projector augmented wave (PAW) [23,24] method as implemented in the Vienna *ab initio* simulation package (VASP) [25,26]. The exchange-correlation effect among electrons was described within the framework of generalized-gradient approximation (GGA), using the functional proposed by Perdew, Burke, and Ernzerhof (PBE) [27]. Note that the semi-core states of W 5$p$ and Cr 3$p$ were treated as valence electrons as well. For the plane wave basis expansion, we used an energy cutoff of 550 eV. We utilized a Monkhorst–Pack **k**-point mesh of 8×8×1 (4×4×1) for the CrX$_3$ (Cr$_{1-x}$W$_x$Cl$_3$) in the structural optimization but a finer **k**-point mesh of 16×16×1 (8×8×1) in calculating the electronic and magnetic properties. The convergence with respect to **k**-point sampling was carefully tested. The vacuum space between adjacent slabs was set to be larger than 15 Å, which is enough to eliminate the spurious image interactions. Both in-plane lattice constants and atomic coordinates were fully relaxed using the conjugate gradient method until the force acting on each atom was less than 0.01 eV/Å. The electron correlation effect for the localized $d$ orbitals of Mn and W atoms was treated by an effective on-site Hubbard term $U$ of 3 eV.

## 3. Magnetic anisotropies and exchange interactions of CrX$_3$ ML

Bulk chromium trihalides have a rhombohedral structure with the R$\bar{3}$ symmetry at low temperature, and a monoclinic structure with the C2/m symmetry at high temperature. After the exfoliation, only one type of ML is found (Fig. 1a). CrX$_3$ MLs consist of halide-Cr-halide triple layers, and Cr atoms are arranged in the honeycomb structure. The unit cell used in the first-principles calculations is shown by the red dashed quadrilateral. The calculated geometric parameters and magnetic properties are summarized in Table 1 and Table S1. It can be clearly seen that lattice parameter increases from CrCl$_3$, CrBr$_3$ to CrI$_3$, which is in accordance with previous reports [15,28,29]. The calculated band structure (Fig. S2a) shows that the single-layer CrCl$_3$ is a FM insulator with an indirect band gap of 2.59 eV.

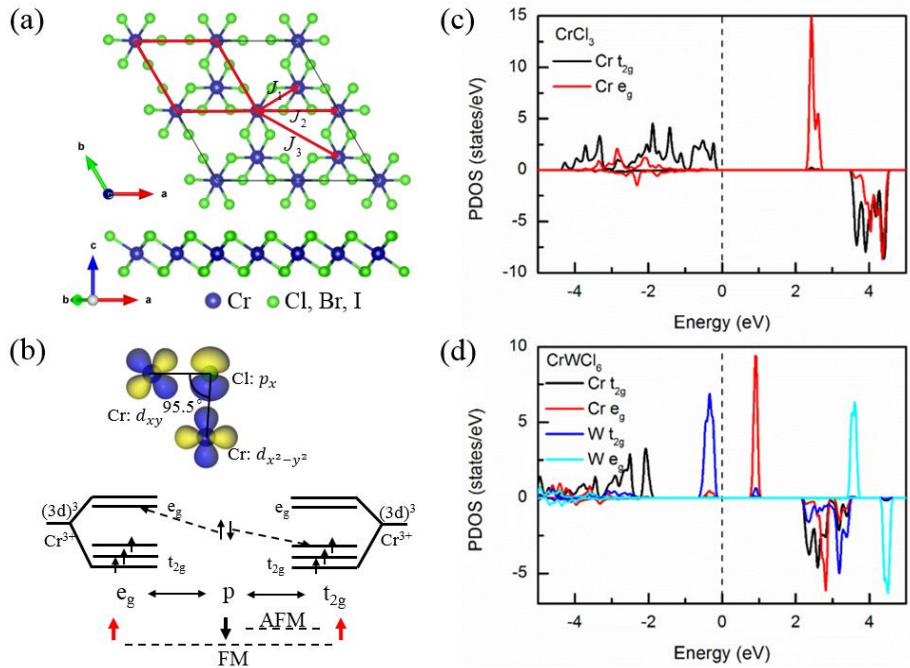

***Figure 1.*** *(a) Top and side views of CrX$_3$ (X = Cl, Br, I) ML. The unit cell is indicated*

by the red dashed quadrilateral. Blue and green balls represent Cr and X atoms, respectively. The first ($J_1$), second ($J_2$) and third ($J_3$) nearest neighboring exchange interaction parameters are indicated by red arrows. (b) Sketch of the FM superexchange interaction in CrCl$_3$ ML via the $d_{xy} - p_{x/y} - d_{x^2-y^2}$ orbitals. (c) The projected density of states (PDOS) of Cr 3d orbitals in CrCl$_3$ ML. (d) The PDOS of Cr and W 3d orbitals in CrWCl$_6$ ML. Positive (negative) values refer to up (down) spins. Fermi level is set at zero energy.

Now we investigate the magnetic anisotropy energy ($E_{MAE}$), a critical parameter to stabilize the long-range magnetic ordering against thermal fluctuations in 2D materials. In principle, the magnetic anisotropy is mainly contributed by two terms: (1) magnetocrystalline anisotropy ($E_{MCA}$) induced by the spin-orbit coupling (SOC), and (2) magnetic shape anisotropy ($E_{MSA}$) resulting from the dipole-dipole interaction, i.e., $E_{MAE} = E_{MCA} + E_{MSA}$. Here, $E_{MCA}$ is obtained by calculating the total energy change (with SOC included) as the magnetization rotates from the in-plane to the perpendicular directions with respect to the CrX$_3$ layer ($E_{MCA} = E_{\parallel}^{SOC} - E_{\perp}^{SOC}$). On the other hand, $E_{MSA}$ ($E_{MSA} = E_{\parallel}^{Dipole} - E_{\perp}^{Dipole}$) is calculated according to the energy of magnetic dipole-dipole interactions as:

$$E^{Dipole} = -\frac{1}{2}\frac{\mu_0}{4\pi r_{ij}^3}\sum_{i=1}^{N}\sum_{j=1}^{r_{max}}\left[\vec{M}_1 \cdot \vec{M}_2 - \frac{3}{r_{ij}^2}\left(\vec{M}_1 \cdot \vec{r}_{ij}\right)\left(\vec{M}_2 \cdot \vec{r}_{ij}\right)\right]. \quad (1)$$

Here $M_i$ represents the local magnetic moments and $r_{ij}$ are vectors that connects the sites $i$ and $j$. Since the this energy converges very slowly with respect to the cutoff of $r_{ij}$ (Fig. S2b), we expanded the range to $r_{max} = 1000$ Å to ensure the numerical reliability.

**Table 1.** The calculated $E_{MCA}$, $E_{MSA}$, $E_{MAE}$, exchange interaction parameters and Curie temperatures of CrX$_3$ and CrWCl$_6$ ML. The units of $E_{MCA}$, $E_{MSA}$ and $E_{MAE}$ are µeV per magnetic atom. Experimental Curie temperatures are also listed for comparison.

| Cases | $E_{MCA}$ | $E_{MSA}$ | $E_{MAE}$ | $J_1$ (meV) | $J_2$ (meV) | $J_3$ (meV) | $T_C$ (K) | Exp. $T_C$ (K) |
|---|---|---|---|---|---|---|---|---|
| CrCl$_3$ | 18 | −58 | -40 | 5.21 | 0.33 | -0.08 | 17* | 27 (bulk)[d] |
| CrBr$_3$ | 157 | −47 | 110 | 4.66 | 0.70 | -0.33 | 31 | 47 (bulk)[d] |
| CrI$_3$ | 655 | −37 | 618 | 8.10 | 1.27 | -0.00 | 70 | 70 (bulk)[d], 45 (1L)[e] |
| CrWCl$_6$ | 1113 | −42 | 1071 | 14.83 | 0.12 | -0.18 | 76 | - |

[d]Ref. [30]

[e]Ref. [8]

*For CrCl$_3$, we assumed an out-of-plane magnetization solely comes from the $E_{MCA}$ term to get a finite value of $T_C$.

The calculated magnetic anisotropy energies of CrX$_3$ MLs are summarized in Table 1. It is interesting that they all have positive $E_{MCA}$, as found in previous theoretical studies. This indicates that their magnetocrystalline anisotropy parts favor the perpendicular magnetization, even though the magnitude of $E_{MCA}$ decreases monotonically as we move up from I, Br to Cl in the periodic table. Note that the monotonic decrease of $E_{MCA}$ is mainly due to the SOC weakening from I, Br to Cl [31]. On the contrary, their shape anisotropy favors an in-plane alignment of magnetization and the magnitude $E_{MSA}$ changes in a narrow range as the local magnetic moments are the same in the three MLs. As a result of the competition, the net magnetic anisotropy energy, $E_{MAE}$, is negative (in-plane easy axis) for CrCl$_3$ ML but is positive (perpendicular easy axis) for CrBr$_3$ or CrI$_3$ MLs, in perfect agreement with experiment. Obviously, $E_{MSA}$ can be comparable to or even larger than $E_{MCA}$, especially when the SOC of chalcogenides is weak. This calls for attention on the consideration of the $E_{MSA}$ for studies of 2D magnetic materials, which has been neglected in most works in this realm [14,15].

Another fundamental parameter for the determination of the Curie temperature of magnetic systems is the exchange interaction. To obtain the exchange parameters, $J_i$, we mapped the DFT total energies of several configurations onto the following classical Heisenberg Hamiltonian:

$$H = -J_1 \sum_{\langle ij \rangle} \vec{S_i} \cdot \vec{S_j} - J_2 \sum_{\langle\langle ij \rangle\rangle} \vec{S_i} \cdot \vec{S_j} - J_3 \sum_{\langle\langle\langle ij \rangle\rangle\rangle} \vec{S_i} \cdot \vec{S_j}, \quad (2)$$

where $J_1$, $J_2$ and $J_3$ represent the nearest, next nearest and third nearest neighbor exchange interactions, respectively (Fig. 1a). For CrX$_3$ MLs, we designed four different magnetic configurations as shown in Fig. S4. As summarized in Table 1, the nearest neighbor exchange interaction ($J_1$) is dominating for the establishment of the FM ground state. The next nearest neighbor interaction ($J_2$) prefers the FM ordering as well, but the weak third nearest neighbor interaction ($J_3$) tends to be antiferromagnetic (AFM). Note that the CrI$_3$ ML has the largest $J_1$ and $J_2$.

The FM $J_1$ originates from the competition between a strong FM superexchange via the near-90° Cr-X-Cr bonds and a weak Cr-Cr AFM direct exchange. Here we take CrCl$_3$ ML as an example. As shown in Fig. 1c, the Cr-d orbitals split into three low-lying half-occupied $t_{2g}$ and two high-lying empty $e_g$ states. The Cr-Cl-Cr angle (95.5°) is close to 90° in the superexchange path (Fig. 1b) and thus gives rise to a FM superexchange interaction according to the Goodenough−Kanamori−Anderson [32-34] rules (Fig. 1b). As the electronic configuration of Cr$^{3+}$ ion is $t_{2g}^3 e_g^0$, the direct exchange between two nearest neighboring Cr$^{3+}$ ions is AFM. However, the direct AFM exchange is weak because of the large distance (3.53 Å) between adjacent Cr ions. As a result, the FM superexchange overtakes the AFM direct exchange, resulting in a net FM $J_1$. As for $J_2$, it involves several possible Cr-Cl-Cl-Cr superexchange paths which contribute either FM or AFM coupling, depending on the corresponding angles [15]. As a result of the competitions between multiple FM and AFM coupling mechanisms, $J_2$ is weak and FM.

## 4. Enhanced Anisotropy and Curie Temperature in CrWCl$_6$

As a strong perpendicular anisotropy [35] is mostly required for 2D magnetic films, it is desirable to find ways to enhance the weak magnetic anisotropy of CrX$_3$ MLs. We perceive that substituting Cr by isovalent W atoms might be a good approach, since W has the same electronic configuration as Cr but much larger SOC. Therefore, we may expect a larger perpendicular magnetic anisotropy and thereby a higher $T_C$. To demonstrate this concept, we used CrCl$_3$ ML as our modeling system. At first, we need to investigate whether Cr and W atoms can be intermixed. We considered 9 different W concentrations by constructing a $2\times 2$ supercell Cr$_{8-x}$W$_x$Cl$_{24}$ with $x = 0 \sim 8$. For each W concentration, there are a large number of possible substitution configurations. In order to find out the optimal distribution patterns of Cr and W atoms, we considered all possible distributions of Cr and W in the $2\times 2$ supercell. The detailed structure configurations for different W concentration ($x$) in the alloyed systems are given in supporting information (Fig. S1). For each ground state of Cr$_{8-x}$W$_x$Cl$_{24}$ (see supporting information), the formation energy $\Delta E_f(\text{Cr}_{8-x}\text{W}_x\text{Cl}_{24})$ is obtained from:

$$\Delta E_f(\text{Cr}_{8-x}\text{W}_x\text{Cl}_{24}) = E(\text{Cr}_{8-x}\text{W}_x\text{Cl}_{24}) - (8-x)E(\text{CrCl}_3) - E(\text{WCl}_3). \qquad (3)$$

The calculated formation energy as a function of W concentration ($x$) is shown in Fig. 2a. We see that the relative energies of the mixed phases are lower by ~0.031 eV per magnetic atom compared to those of the pure phases, namely, CrCl$_3$ and WCl$_3$ MLs. This indicates that the mixed phases can stably exist without phase separation.

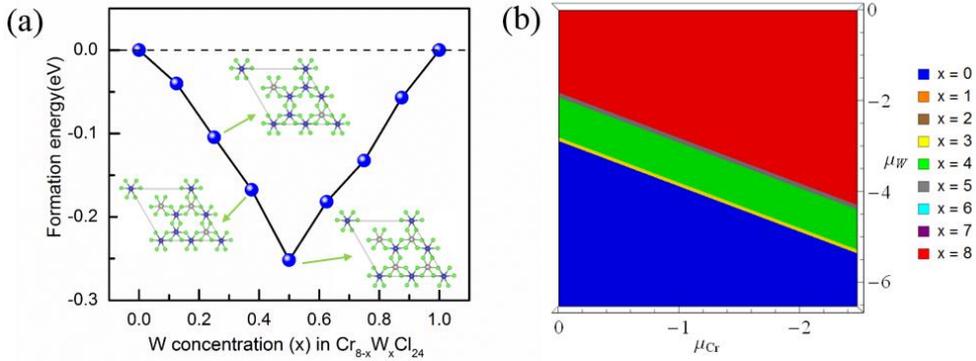

***Figure 2.*** *(a) Formation energy $\Delta E_f(\text{Cr}_{8-x}\text{W}_x\text{Cl}_{24})$ as a function of W concentration (x) in Cr$_{8-x}$W$_x$Cl$_{24}$. Inserts show the lowest-energy configurations for $x = 2 \sim 4$. (b) The phase diagram of the stable region of Cr$_{8-x}$W$_x$Cl$_{24}$ (x = 0~8) with different W concentrations. The number of W atoms in Cr$_{8-x}$W$_x$Cl$_{24}$ ML is labeled by x. The right (bottom) and left (top) boundaries of chromium (tungsten) chemical potential correspond to the so-called Cr (W)-rich (Cr or W bulk as the reservoir) and Cr (W)-poor (Cr or W atom as the reservoir) conditions, respectively.*

To further examine the feasibility of making the (Cr, W) mixed phases in experiments, we calculated the formation enthalpy $\Delta H_f(\text{Cr}_{8-x}\text{W}_x\text{Cl}_{24})$ by varying chemical potentials of Cr, W and Cl as:

$$\Delta H_f\left(\mathrm{Cr}_{8-x}\mathrm{W}_x\mathrm{Cl}_{24}\right) = E\left(\mathrm{Cr}_{8-x}\mathrm{W}_x\mathrm{Cl}_{24}\right) - kT\ln C_8^x - (8-x)\mu_{\mathrm{Cr}} - x\mu_{\mathrm{W}} - 24\mu_{\mathrm{Cl}}. \quad (4)$$

In Eq. (4), $E\left(\mathrm{Cr}_{8-x}\mathrm{W}_x\mathrm{Cl}_{24}\right)$ is the ground state energy of the mixed phase. The second term is the configuration entropy ($T$ = 300 K is used here). $\mu_{\mathrm{Cr}}$, $\mu_{\mathrm{W}}$ and $\mu_{\mathrm{Cl}}$ represent the chemical potentials of Cr, W and Cl relative to the chemical potentials of their elemental phases. Here, we assume that the source of Cl atoms is the same for different W concentrations, so the term, $-24\mu_{\mathrm{Cl}}$, in the Eq. (4) is a constant.

Fig. 2b gives the lowest $\Delta H_f$ of $\mathrm{Cr}_{8-x}\mathrm{W}_x\mathrm{Cl}_{24}$ in the ($\mu_{\mathrm{Cr}}$, $\mu_{\mathrm{W}}$) plane within ranges of $-2.48\,\mathrm{eV} < \mu_{\mathrm{Cr}} < 0$ and $-6.54\,\mathrm{eV} < \mu_{\mathrm{W}} < 0$. One may see that the mixed states can survive in a large region for $x = 4 \sim 6$, which means that mixing intermediate concentration of W into CrCl$_3$ ML is feasible in the W-rich condition around $\mu_W = -4$ eV. The stable region in this phase diagram has no obvious changes with decreasing the value of $U$ in DFT calculations (see Fig. S4 in supporting information), and the evenly mixed Cr$_4$W$_4$Cl$_{24}$ with $x$ = 4 (i.e. CrWCl$_6$) appears to be the most preferential configuration for all cases.

As the Cr and W atoms tend to rather uniformly, the unit cell can be reduced back to CrWCl$_6$ ML, which makes the discussion and comparison rather straightforward. Because of the strong SOC of W, $E_{\mathrm{MCA}}$ of the CrWCl$_6$ ML has a large positive value, 1114 μeV per magnetic atom. As the value of $E_{\mathrm{MSA}}$ is almost unchanged from that of the Cr$_2$Cl$_6$ ML, the magnetic easy axis of CrWCl$_6$ ML is stably perpendicular as desired, with a giant MAE of 1071 μeV per magnetic atom (Table 1). This value is even comparable with MAEs of many FM transition metal thin films such as Fe, Co, and Ni on different substrates [36].

To investigate the physical origin of the giant perpendicular $E_{\mathrm{MCA}}$ of CrWCl$_6$ ML, we plotted its band structure in Fig. 3a. After W atoms are introduced, the band gap is 1.07 eV, with the unoccupied spin up states of the Cr$_2$Cl$_6$ ML moving toward the Fermi level. However, we can't draw a simple conclusion for the large $E_{\mathrm{MCA}}$ solely from the reduced band gap. Here, we analyze the tendency of $E_{\mathrm{MCA}}$ of the CrWCl$_6$ ML with respect to the shift of the Fermi level, using the torque method and rigid band model [37,38]. Note that the $E_{\mathrm{MCA}}$ obtained from the toque method, 1089 μeV per magnetic atom is very close to that listed in Table 1 (1114 μeV), indicating the reliability of the torque method for these systems. We decompose $E_{\mathrm{MCA}}$ into contributions from different atoms or spin channels and show the results in Fig. 3b. It can be seen that W atoms contribute the most to $E_{\mathrm{MCA}}$, i.e., 990 μeV per magnetic atom. From the spin resolutions, we find that the dominating part is $\Delta E^{\uparrow\downarrow/\downarrow\uparrow}$ (Fig. S5b), which contributes 60% of $E_{\mathrm{MCA}}$. If we move the Fermi energy down by 0.5 eV, $\Delta E^{\uparrow\downarrow/\downarrow\uparrow}$ exhibits a sudden change as shown in Fig. S5b, indicating that electronic states near -0.5 eV play the main role in producing the large positive $E_{\mathrm{MCA}}$ of the CrWCl$_6$ ML. By projecting wave functions of states around this energy to the W site (Fig. 3a), we reach a conclusion that the perpendicular $E_{\mathrm{MCA}}$ mainly results from the SOC interactions between the occupied spin-up $d_{xz/yz}$ (bands -1, -3 in Fig. 3a) and unoccupied spin-down $d_{z^2}$ states (band 1 in Fig. 3a) of W through the $L_x$ operator. As W has a large SOC strength, this coupling produces a large perpendicular $E_{\mathrm{MCA}}$.

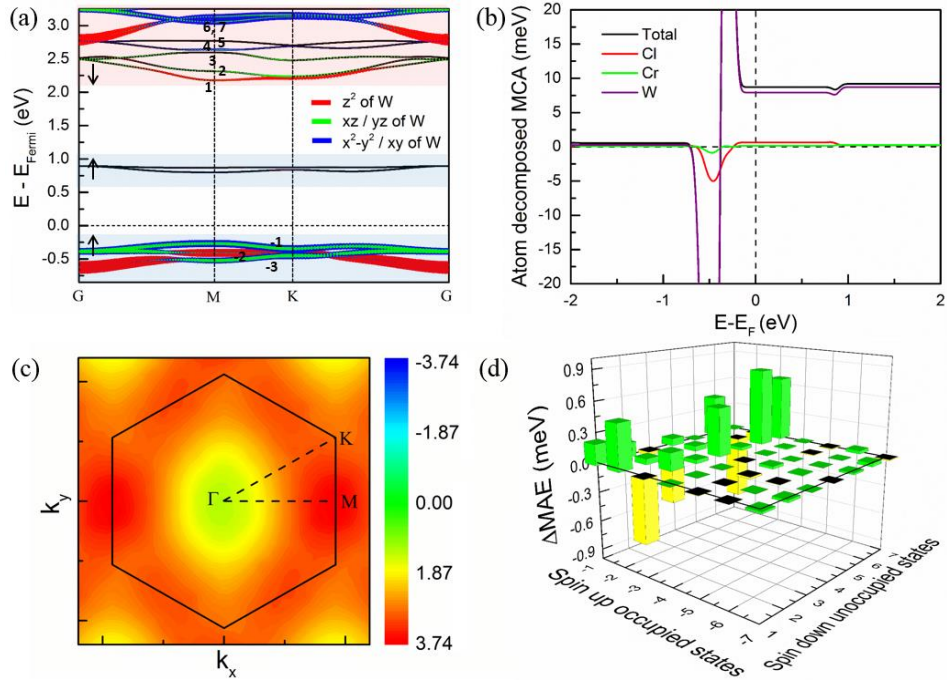

***Figure 3.*** *(a) Orbital resolved band structure of $CrWCl_6$ ML. The $|m| = 0, 1, 2$ orbitals of W are represented by red, green, blue lines, respectively, and the line width scales with their weights. Band indexes are marked by numbers. The up and down spin states are indicated by the light blue and red shades, respectively. (b) The decomposition of the total $E_{MCA}$ to the contributions of different atoms in $CrWCl_6$ ML. The black line shows the total $E_{MCA}$, and the red, green and purple lines represent the contributions from Cl, Cr and W atoms, respectively. (c) Distribution of $\Delta E_{MCA}$ in the BZ. The blue and red colors represent negative and positive contributions, respectively. (d) Contributions of each pair of spin-up occupied states and spin-down unoccupied states for summation over all the K and M points in the first BZ.*

We note that the nearest neighbor exchange coupling parameter, $J_1$, is also large for $CrWCl_6$ ML, up to 14.8 meV (Table 1), due to the stronger W-Cl hybridization. The energy separations between the occupied W-$t_{2g}$ orbitals and empty Cr-$e_g$ orbitals in $CrWCl_6$ are largely reduced compared to their counterparts in $CrCl_3$ [39] (Fig. 1c and Fig. 1d). Since the FM superexchange arises from the virtual exchange between $t_{2g} - p - e_g$ orbitals, this reduction in energy separation enhances the FM superexchange interaction, and thus leads to a larger $J_1$.

Interestingly, the perpendicular magnetic anisotropy of $CrWCl_6$ ML can be further enhanced by strains. For example, an in-plane tensile strain of 5% may increase $E_{MCA}$ of the $CrWCl_6$ ML to 2375 μeV per magnetic atom, more than twice as large as the unstrained one. By decomposing $\Delta E_{MCA} = E_{MCA}^{strained} - E_{MCA}^{unstrained}$ to the contributions of different **k**-points in the first Brillouin zone (BZ) (Fig. 3c), we found that $\Delta E_{MCA}$ mainly occurs around the M point. We further resolve the $\Delta E_{MCA}$ to contributions of different pairs of occupied-unoccupied states according to the second-order perturbation formula (see Part IV in supporting information). The results are shown in Fig. 3d, where the band indices are given according to Fig. 3a. Clearly, the eye-catching pairs are (-1, 2), (-3, 4) and (-3, 7). Fig. S5 gives more details about their energy separation and matrix elements.

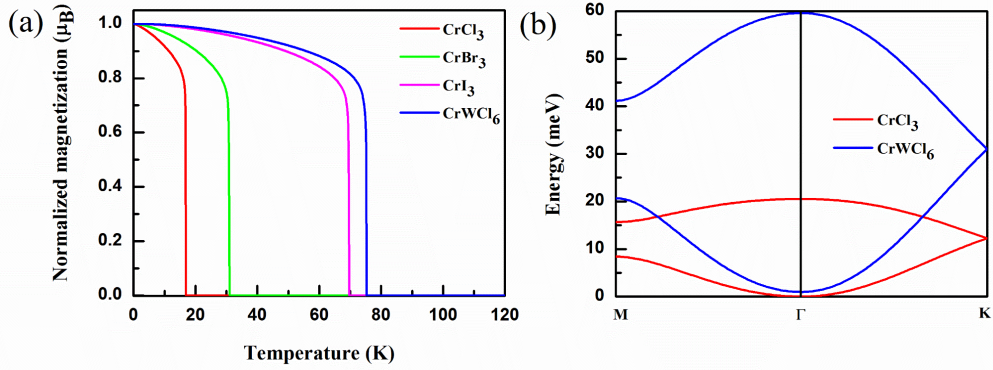

**Figure 4.** *(a) The renormalized magnetization $\frac{M(T)}{M_0}$ as a function of temperature T. $T_C$ is determined by the temperature at $M(T)=0$. (b) Spin-wave excitation (magnon) spectrums of CrCl$_3$ ML and CrWCl$_6$ M.*

Finally, we determined the $T_C$ of CrWCl$_6$ using the renormalized spin-wave theory (RSWT) [40,41] to show the benefit of Cr-W intermixing. It is recognized that the linear spin-wave theory (LSWT) typically overestimates $T_C$ of 2D magnetic films [40,41], and the inclusion of multi-magnon scattering (i.e., the non-linear corrections in [31]) is essential to get reliable estimation of their $T_C$. Through the studies of magnetic field dependence of $T_C$ in few-layer and bulk Cr$_2$Ge$_2$Te$_6$, it was shown that the RSWT method can reach quantitative agreement with experiments [7]. Here, we outline the essence of the RSWT. In LSWT, through the Holstein-Primakoff transformation, the spin operator $\mathbf{S}_{l\nu} = (S^x_{l\nu}, S^y_{l\nu}, S^z_{l\nu})$ on site $\nu$ in the **l**-th unit cell in Eq. (2) is rewritten by the deviation creation and annihilation operators $a^+_{l\nu}$ and $a_{l\nu}$, i.e., $S^+_{l\nu} \approx \sqrt{2S} a_{l\nu}$, $S^-_{l\nu} \approx \sqrt{2S} a^+_{l\nu}$, and $S^z_{l\nu} \approx S - a^+_{l\nu} a_{l\nu}$. In the RSWT, however, $\mathbf{S}_{l\nu}$ is expanded up to second order terms of $a^+_{l\nu}$ and $a_{l\nu}$ to take into account the magnon-magnon interaction at finite temperatures [41]. In this case, the operator $S^z_{l\nu}$ is the same as that in LSWT but spin ladder operators $S^+_{l\nu}$ and $S^-_{l\nu}$ are in the form of

$$S^+_{l\nu} \approx \sqrt{2S}\left(a_{l\nu} - a^+_{l\nu} a_{l\nu} a_{l\nu}/4S\right), \qquad (5)$$

$$S^-_{l\nu} \approx \sqrt{2S}\left(a_{l\nu}{}^+ - a^+_{l\nu} a^+_{l\nu} a_{l\nu}/4S\right). \qquad (6)$$

Using Eq. (5-6) and the Fourier transformation $a^+_{l\nu} = \frac{1}{\sqrt{N}}\sum_{\mathbf{k}} e^{-i\mathbf{k}\cdot\mathbf{b}} b^+_{\mathbf{b}\nu}$ and $a_{l\nu} = \frac{1}{\sqrt{N}}\sum_{\mathbf{k}} e^{i\mathbf{k}\cdot\mathbf{b}} b_{\mathbf{b}\nu}$, we transform the spin Hamiltonian Eq. (2) from the real space to the reciprocal space and the resulting Hamiltonian contains the magnon-magnon interaction [41]. Following the Bose-Einstein statistics, the total magnetization $M(T)$ as a function of temperature $T$ is expressed as:

$$\frac{M(T)}{M_0} = 1 - \frac{1}{nNS}\sum_{\pm}\sum_{\mathbf{k}}\left[\exp\left(\frac{\hbar\omega_{\mathbf{k}\pm}}{k_B T}\right) - 1\right]^{-1} \quad (7)$$

where $N$ and $n$ are the numbers of unit cells and magnetic basis atoms in each unit cell, respectively. $M_0$ denotes the fully polarized magnetization at $T = 0$ K. "$\pm$" represents the two magnon braches, i.e., the in-phase acoustic mode and the anti-phase optical mode. $\omega_{\mathbf{k}\pm}$ is the magnon spectrum solved from Eq. (6). The Curie temperature, $T_C$, is determined by the temperature at which the total magnetization becomes zero, i.e. $M(T) = 0$.

The calculated total magnetizations of $CrCl_3$, $CrBr_3$, $CrI_3$ and $CrWCl_6$ MLs by RSWT as a function of temperature are shown in Fig. 4a, and their values of $T_C$ are listed in Table 1. It should be pointed out that, in principle, $CrCl_3$ ML cannot sustain a long-range magnetic order since its magnetic easy axis is in-plane, as mentioned previously. For a better comparison, we ignore the $E_{MSA}$ term for the $CrCl_3$ ML and assume an out-of-plane magnetization that solely comes from $E_{MCA}$. Nevertheless, this hypothetical $CrCl_3$ ML still has a low $T_C$, only 17 K. By mixing with W, the $T_C$ of $CrWCl_6$ ML can be as high as 76 K, even higher than that of $CrI_3$ ML. From the spin wave (magnon) spectrum as shown in Fig. 4b, we see that the optical branch of $CrWCl_6$ ML shifts to a much higher energy compared with that of $CrCl_3$ ML, indicating the difficulty to excite it and hence the $T_C$ is significantly improved.

## 5. Summary and conclusions

In conclusion, we systematically investigated the 2D ferromagnetism of $CrX_3$ and $Cr_{8-x}W_xCl_{24}$ ($x = 0\sim8$) MLs. We found that the magnetic shape anisotropy is responsible for the experimentally observed in-plane magnetic easy axis of $CrCl_3$ ML, indicating that it is vital to consider this contribution for studies of weak magnetic anisotropy thin films. To enhance the magnetic anisotropy and $T_C$ of the $CrCl_3$ ML, we propose to substitute some Cr atoms with the isovalent W atoms. The theoretical feasibility and experimental maneuverability of the intermixed $Cr_{8-x}W_xCl_{24}$ MLs were thoroughly studied. We find that the intermediate concentration doping with $x = 4\sim6$ is energetically preferred and thereby can be easily synthesized in experiments. As desired, the $CrWCl_6$ ML has a significantly enhanced perpendicular magnetic anisotropy, a large exchange interaction, and hence a high Curie temperature up to 76 K. Our work gives insights into understanding the two-dimensional ferromagnetism of van der Waals magnetic MLs and suggests a route toward improving the performance of $CrX_3$ MLs for spintronic applications.

# Acknowledgement

Work was supported by DOE-BES (Grant No. DE-FG02-05ER46237). Computer simulations were partially performed at the U.S. Department of Energy Supercomputer Facility (NERSC). F.X and Z.W. acknowledge support by the Basic Research Program of China under Grant No. 2015CB921400.